\documentstyle[prl,aps,psfig]{revtex}  
\begin{document}
\twocolumn

\title{Large-scale superconductivity-induced conductance suppression
in mesoscopic normal-superconducting structures. }

\author{R. Seviour$^{*}$, C.J. Lambert$^{*}$ and M. Leadbeater$^+$}
\address{
* School of Physics and Chemistry,
Lancaster University, Lancaster LA1 4YB, U.K.
\\ + Max-Planck-Institut f\"{u}r Physik komplexer Systeme,
N\"{o}thnitzer Strasse 38,
01187 Dresden,
Germany.}
\date{\today}
\maketitle

\begin{abstract}
Experiments on hybrid superconducting normal-metal structures have revealed that
even in the absence of tunnel junctions the onset of superconductivity
can lead to a decrease in the electrical conductance by
an amount many orders of magnitude greater than $e^2/h$.
In this Letter we provide a theory of this phenomenon
which shows that it originates from an instability in
4 - probe conductance measurements which is absent
from 2-probe measurements.
We compare the zero-bias, zero-temperature
4-probe conductances $G_{N}$ and $G_{S}$
of a normal diffusive metal
in contact with a superconductor in both the normal (N) and superconducting (S)
states respectively. 
In the absence of
tunnel barriers,  the ensemble average of the
difference $\delta G=G_{S}-G_N$ vanishes, in agreement with
quasi-classical theory. However we also predict that there exists macroscopic
sample specific fluctuations in $\delta G$, which lie beyond quasi-classical
theory and allow large negative values of $\delta G$ to occur.

\end{abstract}

During the past few years, studies of the subgap conductance
of  hybrid normal-superconductor (N-S) structures have led to the identification
of a small number of paradigms of phase-coherent transport.
The earliest of these is the zero bias anomaly observed in
normal-insulator-superconductor (N-I-S)
structures \cite{c1} and the related finite bias anomaly found
in N-S structures with clean interfaces. At high temperatures $T>T^*$
and bias-voltages $V>V^*$, where for
a N-metal of length $L$ and diffusion coefficient $D$,
 $k_BT^*=eV^*=\sqrt(D/L^2)$, the conductances of both types of structure
vary as $1/\sqrt T$ and  $1/\sqrt V$. For a clean interface there also exists a
conductance maximum at $V^*$, $T^*$ and therefore at low-temperature
and voltage a re-entrance to the low-conductance state occurs \cite{c2,c3} .
This re-entrance phenomenon is also observed in a second paradigm
of phase-coherent transport, which arises when a normal metal is in
contact with two superconductors, whose order parameter phase
difference $\phi$ can be varied by some external means. In this case the
conductance is an oscillatory function of $\phi$, with an amplitude
which typically exhibits a maximum at $T^*$ and $V^*$  
\cite{c2,c4,c5,c6,c7,c8,c9,c10}. Unlike the Josephson current which 
decays exponentially with $T/T^*$, such conductance oscillations decay
only as a power-law. A third signature of phase-coherent transport is the 
long-range nature of these effects, which typically decay as a power-law in 
$L/L^*$, where $L^*=\sqrt(D/eV)$. This behaviour is in sharp contrast
with the exponential decay of the Josephson effect and has been observed in
a number of experiments \cite{c11,c12}.
 
All of the above phenomena can be explained by quasi-classical theory
and to large extent are well-understood. There are however two further
phenomena which are as yet unexplained by current theories of
 phase-coherent transport, namely the macroscopic suppression of electrical
 conductance by superconductivity \cite{c12,c13}
  and re-entrant paramagnetism in
superconducting wires coated with normal metal \cite{c14}. Both of these
effects have resisted detailed explanation, but must be understood if a
complete theory of phase coherent transport is ever to be claimed.
 In this Letter we present a theory of the first of these phenomenon, which
 has been observed in several experiments \cite{c12,c13},
 involving both non-magnetic
 (ie Silver) and magnetic (ie Nickel) N-components, in contact with a
 superconductor via a clean interface. In contrast with
 quasi-classical theories \cite{c15,c16} which
 predict that when the superconductor is in its
 normal state, the zero temperature, zero-bias conductance $G_N$ is identical
to the conductance $G_{S}$ in the superconducting state, these
 experiments show that the difference $\delta G=G_N-G_{S}$ can be of
order $G_N$ and with a sample-specific sign.
Since tunnel barriers and grain boundaries are absent
from such samples, this effect is an embarassment to current
theory. At finite bias,
quasi-classical theory \cite{c16} has shown that a negative $\delta G$ 
is possible, but the effect discussed in [16]
is too small to account for the large 
changes found experimentally. In this Letter we suggest an explanation of 
this phenomenon which lies outside quasi-classical theory.

To go beyond quasi-classical theory,
we adopt a general scattering approach to dc transport,
which was initially developed to describe phase-coherent transport in dirty
mesoscopic superconductors \cite{c17} and is based on fundamental current-
voltage relations derived in \cite{c18,c19}. For simplicity in this Letter,
 we focus solely on the zero-voltage, zero-temperature
  conductance, for the structure
 shown in figure 1(a), which captures the essential physics of
 the experimental setup of reference
 \cite{c13} and comprises a disordered superconductor of
 condensate potential $\mu$, sandwiched between two pieces of diffusive 
 normal metal. The latter is in contact with four normal reservoirs at voltages
$v_j, j=1,\dots ,4$. Leads 1 and 2 carry a current $I$, while leads 3 and 4
carry no current and therefore form the voltage probes of a 4-probe
conductance measurement.  As in the experiments,
the superconductor is chosen to be longer
than the superconducting coherence length, so that sub-gap quasi-particles
 are not transmitted through the superconducting region.

 In the linear-response limit, at zero temperature, the conductance of a
 phase-coherent
 structure may be calculated from the fundamental current voltage relationship
\cite{c18,c19},
\begin{equation} 
I_i=\sum_{j=1}^4a_{ij}( v_j - v),
\label{a4}
\end{equation}

The above expression relates the current $ I_i$ 
from a normal reseviour $i$ to the voltage differences $(v_j - v)$, 
where $v=\mu/e$
and the sum is over the 4 normal leads connected to the scattering region.
The $a_{ij}$'s are linear conbinations of the normal and Andreev scattering
coefficients and in the absence of superconductivity satisfy $\sum_{j=1}^La_{ij}
=\sum_{i=1}^La_{ij} = 0 $ in which case the left hand side of equation \ref{a4}
becomes independent of $v$. In units of
$2e^2/h$ \cite{c18,c19}, $a_{ii}=N_i+R_i^A-R_i^O$ and $a_{ij\ne i}=T_{ij}^A-T_{ij}^O$,
where $T_{ij}^A$, $T_{ij}^O$ are Andreev and normal transmission coefficents from
probe $j$ to probe $i$, $R_i^A$, $R_i^O$ are Andreev and normal refelction
coefficients from probe $i$ and $N_i$ is the number of open scattering
channels  in lead $i$.

As an aid to comparing transport coefficients in the
normal and superconducting states, it is convenient to eliminate $v$
from equation \ref{a4} by the following procedure suggested in  \cite{c20,c21},
which yields

\begin{equation} 
I_i=-\sum_{j=1}^4G_{ij} v_j
\label{ce2}
\end{equation}

where
\begin{equation} 
G_{ij} = -a_{ij} + x_i y_j / s
\label{ce3}
\end{equation}

\begin{equation} 
x_i=\sum_{j=1}^4a_{ij}
\label{ce4}
\end{equation}
\begin{equation}
y_j=\sum_{i=1}^4a_{ij}
\label{ce5}
\end{equation}
\begin{equation}
s=\sum_{i=1}^4x_i = \sum_{j=1}^4y_j
\label{ce6}
\end{equation}

Equation \ref{ce2} is a fully quantum mechanical current - voltage relation 
which allows us to represent a multiprobe structure by an equivalent circuit,
which for 4 probes is shown in figure 1 (b). In contrast with the normal case
however, the "conductances" $G_{ij} (i\neq j)$ are not necessarily positive.
Furthermore, as noted in \cite{c20,c21}
$G_{ij} (H) = G_{ji} (H^*)$, where $H$ is the Hamiltonian of the system
and therefore in the presence of a magnetic field,  $G_{ij}\neq G_{ji}$ .

Setting $I_1 = -I_2 = I, I_3=I_4=0$ and solving 
equation \ref{ce2} for the 4 probe conductance yields \cite{c20}

\begin{equation} 
G= \frac{d}{(G_{42}G_{31} - G_{41}G_{32} )}
\label{ce7}
\end{equation}
where $d \geq 0$ is the determinant of the $3{\rm x}3$ matrix obtained by removing
the third row and column from the matrix $G_{ij}$
\footnote{ As a simple example, for a symmetric structure where
 $G_{13}$ = $G_{24}$
$G_{12}=G_{34}$ and $G_{14}=G_{23}$ equation \ref{ce7} reduces to
$$
G_{i}=\frac{2(G_{43}+G_{42})(G_{43}+G_{41})}{G_{42}-G_{41}}
$$}.
 In
what follows, we consider only the case where $H$ is real (i.e. a real order 
parameter and no magnetic field), so that $G_{ij}=G_{ji}$.

As noted in \cite{c20} in the presence of disorder, the various transmission 
and reflection coeffcients can be computed by solving the
Bogoliubov - de Gennes equation on a tight-binding lattice of
sites, each labelled by an index $i$ and possessing
a particle (hole) degree of freedom $\psi(i)$ $(\phi(i))$.
In the presence of local s-wave pairing described by a
superconducting order parameter $\Delta_i$, this takes the form

\begin{eqnarray}
\begin{array}{c c}
E\psi_i
=&\epsilon_i \psi_{i}
-\sum_{\delta} \gamma \left(  \psi_{i+\delta} + \psi_{i-\delta} \right)
+ \left( \Delta_{i} \phi_{i}\right)\\
E\phi_i =&-
\epsilon_i \phi_{i}
+\sum_{\delta} \gamma  \left(  \phi_{i+\delta}+ \phi_{i-\delta}\right)
+\Delta^*_{i}\psi_{i},\\
\end{array}
\label{2}
\end{eqnarray}

The Hamiltonian of eq.(\ref{2}) has been used to describe all of the
paradigms [1-11].
 In the normal diffusive region 
of figure 1(a), the on-site energy $\epsilon_i$ is chosen to be a random number,
uniformly distributed over the interval $\epsilon_0 -W/2$ to $\epsilon_0+W/2$, 
whereas in the clean N-regions $\epsilon_i=\epsilon_0$. In the S-region, the 
order parameter is set to a constant, $\Delta_i=\Delta_0$, while in all other 
regions, $\Delta_i=0$. The nearest neighbour hopping element $\gamma$ merely 
fixes the energy scale (ie the band-width), whereas $\epsilon_0$ determines the
band-filling. In what follows we choose $\gamma=1$.
By numerically solving for the scattering matrix of equation 8,
exact results for the dc conductance can be
obtained and therefore if large negative values of $\delta G$ were not 
obtainable from such an approach, one would have to conclude that the 
experiments of \cite{c13} are not describable by mean-field BCS theory. 
In what follows, we employ an exact recursive Greens function technique to
compute the scattering matrix of the structure of figure 1a. To obtain results
which are typical of a normal metal, large-scale simulations are carried out,
in which the sample dimensions are two orders of magnitude larger than the
Fermi wavelength.

Figure 2(a) shows results for ten different disorder realizations 
both in the normal state ($\Delta_0=0$) and in the superconducting
state ($\Delta_0=0.1$). As is seen from figure 2(a), for certain
disorder realizations the normal state 
conductance $G_N$ is much greater than the conductance $G_S$ 
in the superconducting state, yielding a large negative value for
$\delta G=G_S -G_N $.
In contrast the ensemble average of $\delta G$, obtained by
ensemble averaging over 500 disorder realisations,
 is negligible compared to $G_N$,
in agreement with quasiclassical theory. As a prelude to the above simulations,
to ensure that the disordered metal is diffusive, the conductance $G_N$ was 
calculated as a function of length $L$ with $\Delta_0$ set to zero. In the 
diffusive region, the mean free path $l$ is independent of length and in units 
of the lattice spacing is given by $l= (h/2e^2)G_NL/M$, where $M$ is
the sample width. For the system used in
figure 2(a), $L=M= 200$ sites, $\epsilon_0=0.2$ $W=2.0$ and $l\approx 4.5$.

The fact that large negative values of $\delta G$ are describable by
equation \ref{ce7} is a central result of this Letter. By examining
individual scattering coefficients, we now argue that this effect
arises from the occurrence of small denominators in equation \ref{ce7}.
Indeed, in the normal state, if by chance $G_{42}G_{31}
 \approx G_{41}G_{32}$
then $G_N$ will be anomalously large.
In this case, if the onset of superconductivity causes the denominator
to increase, then a large negative $\delta G$ can occur.

Figure 2(c) shows the products $G_{42}G_{31}$ and
 $G_{41}G_{32}$ for each of the normal state samples in figure 2(a). As an example,
figure 2(a) shows that for sample no. 5,
 the normal state conductance is relatively large,
while figure 2(c) shows that
$G_{42}G_{31}$ and $G_{41}G_{32}$ are approximately equal.
Figure 2(b) shows the corresponding conductances
 when the samples are superconducting.
As one might intuitively expect, the onset of superconductivity causes
an increase in $G_{42}G_{31}$. Nevertheless
the associated increase in the denominator of equation \ref{ce7}
leads to a large-scale decrease in the 4-probe conductance $G$.

The aim of this Letter is to demonstrate that
the phenomenon of large-scale superconductivity-induced conductance suppression
in metallic mesoscopic structures can be understood through an exact solution
of the Bogoliubov - de Gennes equation.
The results of figure 2
demonstrate that for specific disorder realizations, as a consequence
of an instability due to a small denominator in equation
\ref{ce7},  the onset of
superconductivity can decrease the total conductance, even though the 
individual conductances ($G_{ij}$) may increase and even though the 
fluctuations in the $G_{ij}$'s are merely of order $e^2/h$.
Since the
ensemble averaged total conductances for both the normal and superconducting
state are equal, this phenomenon is not contained in  a quasi-classical
description.

In addition to the calculations leading to figures 2,
we have carried out
extensive numerical simulations of  two-probe
structures and  in no-case have we
found large negative conductance changes in the diffusive limit. 
Figure 3 shows results for 10 disorder realizations obtained for a N-S
structure of width M = 150 sites comprising a semi-infinte crystalline N-lead, 
in contact with a N-diffusive region of length L = 30, which
in turn makes contact via a clean interface to a  semi-infinte crystalline 
S-lead. In all cases $\delta G$ is negative and of order $e^2 / h$. A detailed 
analysis of the Lyapunov exponents shows that this is a consequence of the weak 
localisation effect predicted in \cite{c22}, which is too small to account
for the large-scale effect described in this Letter.

We believe that the results of figure 2 capture
the essential physics of
the experiments of \cite{c13}, although measured values for
the conductances $G_{ij}$ are not available, thereby making
a quantitative comparison impossible.
The crucial question is whether or not the condition
$G_{42}G_{31}  \approx G_{41}G_{32}$ is satisfied in the normal-state.
Although the experimental samples are somewhat elongated, this condition
may arise as a consequence of additional disorder at the ends
of the sample, where contact with the leads is made.
For the future, it would be of interest to carry out
an experiment on shorter samples, with a length:width ratio less than or of order
unity,
since the denominator
of equation \ref{ce7} will be small.
The sign of $\delta G$ will be sample dependent and therefore by switching on 
a magnetic field,  fluctuations in $G_{ij}$ of order $e^2/h$
should generate
both positive and negative macroscopic fluctuations in  $\delta G$.
Indeed there is no restriction on the sign of $G_N$ and $G_S$ and therefore
such an experiment would also illustrate the prediction of superconductivity-induced
negative 4-probe conductances discribed in reference \cite{c20}.

\begin{figure}
\centerline{\psfig{figure=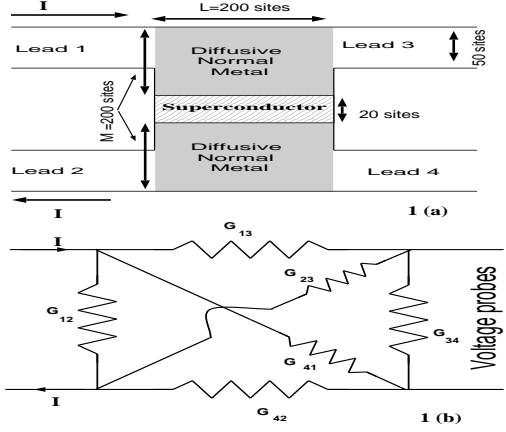,width=7 cm,height=6cm}}
\caption{\tiny Fig 1(a) The N-S-N structure analyzed numerically. Fig 1(b)
The equivalent quantum resistor network representing the structure shown in
figure 1 (a), where the $G_{ij}$ are given by equation 3.}
\label{fig1}
\end{figure}

\begin{figure}
\centerline{\psfig{figure=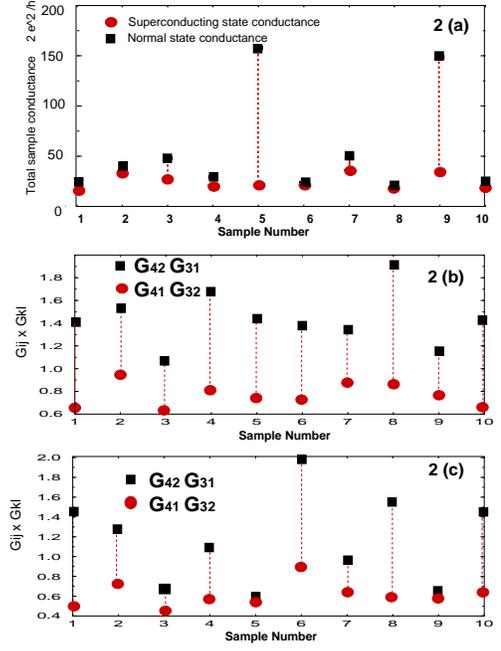,width=7cm,height=9cm}}
\caption{\tiny
2(a) Results for the electrical conductance $G_{N}$,$G_{S}$ of the
structure shown in figure 1a, for 10
samples of differing disorder realizations.
 2(b) Results for the products
$G_{41}G_{32}$ and $G_{42}G_{31}$ for each of
 the 10 samples from figure 2(a), when the system is in the superconducting
 state.
 2(c) Results for the products
$G_{41}G_{32}$ and $G_{42}G_{31}$ for each of
 the 10 samples from figure 2(a), when the system is in the normal
 state.}
\label{fig2}
\end{figure}

\begin{figure}
\centerline{\psfig{figure=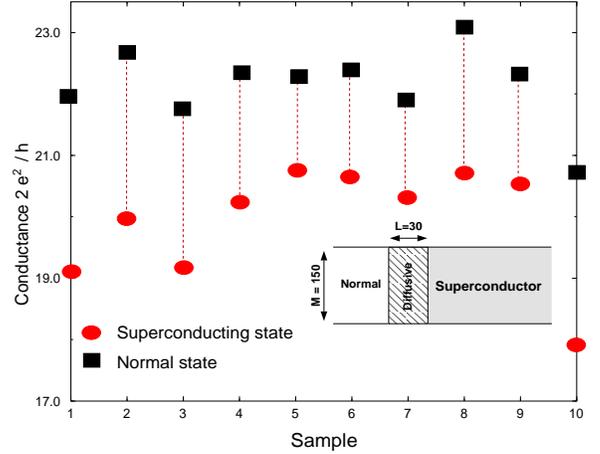,width=8 cm,height=10cm}}
\caption{\tiny  Numerical results for 10 disorder realisations, obtained for
the N-S geometry shown in the insert.} 
\label{fig3}
\end{figure}

\end{document}